\documentclass[numberedappendix,apj]{emulateapj}

\usepackage{amsmath,color}

\newcommand{\bl}{}
\newcommand{\rd}{}

\shorttitle{Dust dams in `transition' discs}
\shortauthors{}

\begin{document}


\title{Accreting planets as dust dams in `transition' discs.}

\author{James E. Owen}
\affil{Canadian Institute for Theoretical Astrophysics, 60 St. George St., Toronto, M5S 3H8, Canada}
\email{jowen@cita.utoronto.ca}



\begin{abstract}
We investigate under what circumstances an embedded planet in a protoplanetary disc may sculpt the dust distribution such that it observationally presents as a `transition' disc.  We concern ourselves with `transition' discs that have large holes ($\gtrsim 10$ AU) and high accretion rates ($\sim 10^{-9}-10^{-8}$ M$_\odot$ yr$^{-1}$). Particularly, those discs which photoevaporative models struggle to explain. Assuming the standard picture for how massive planets sculpt their parent discs, along with the observed accretion rates in `transition' discs, we find that the accretion luminosity from the forming planet is significant, and can dominate over the stellar luminosity at the gap edge. This planetary accretion luminosity can apply a significant radiation pressure to small ($s\lesssim 1\mu$m) dust particles provided they are suitably decoupled from the gas. Secular evolution calculations that account for the evolution of the gas and dust components in a disc with an embedded, accreting planet, show that only with the addition of the radiation pressure can we explain the full observed characteristics of a `transition' disc (NIR dip in the SED, mm cavity and high accretion rate). At suitably high planet masses ($\gtrsim 3-4$ M$_J$), radiation pressure from the accreting planet is able to hold back the small dust particles, producing a heavily dust-depleted inner disc that is optically thin (vertically and radially) to Infra-Red radiation. We use our models to calculate synthetic observations and present a observational evolutionary scenario for a forming planet, sculpting its parent disc. The planet-disc system will present as a `transition' disc with a dip in the SED, only when the planet mass and planetary accretion rate is high enough. At other times it will present as a disc with a primordial SED, but with a cavity in the mm, as observed in a handful of protoplanetary discs. 


\end{abstract}
\section{Introduction}
Where, when and how planets form is a currently unsolved problem in astrophysics. Central to this issue is how planets interact and sculpt the environment in which they are forming. The large number of observed exoplanets indicates that protoplanetary discs which churn out planets are the norm, rather than a rare occurrence. However, connecting the properties of protoplanetary discs and exoplanets with a coherent picture of planet formation and evolution remains elusive. 

Protoplanetary discs are made up of gas and dust particles; while the dust is only a minor component by mass ($\sim 1\%$) it dominates the opacity of the disc material. Therefore, most observational indicators of protoplanetary disc evolution probe the evolution of the dust component, rather than the gas which drives the dynamics. Protoplanetary discs appear to live for $\sim 1-10$ Myr \citep{haisch01,hernandez07}, until they are destroyed, most likely by photoevaporation \citep{clarke01,alexander06,owen11,owen12,alexander14}. During their lifetime, discs are primarily optically thick out to large radius ($\gg 10$ AU), with accretion rates that decline with time in the range $10^{-6}-10^{-10}$ M$_\odot$ yr$^{-1}$ \citep[e.g.][]{hartmann98}, and masses in the range $10^{-3}-10^{-1}$ M$_\odot$ \citep[e.g.][]{andrews05,andrews07}.

However, a small fraction of protoplanetary discs show a lack of opacity at NIR wavelengths, but return to levels comparable with primordial discs at MIR wavelengths \citep{strom89,skrutskie90}. The rarity of this population of protoplanetary discs, coupled with the fact that their inner discs appeared to be cleared of dust, has led many authors to suggest that these discs may be caught in the act of evolving from a young star with a primordial disc to a disc-less star, and have been aptly named `transition' discs \citep[e.g.][]{kenyon95,ercolano11,koepferl13,espaillat14}. Many mechanisms have been proposed in order to explain the properties of `transition' discs: tidal truncation by a companion \citep[e.g.][]{calvet05,rice06,dodson11}, photoevaporation \citep[e.g.][]{clarke01,owen11}; grain-growth \citep[e.g.][]{dullemond05}; photophoresis \citep{krauss07}; MRI driven winds \citep{suzuki09} and magnetic winds \citep{armitage13}. In specific, individual cases these models can reproduce the observed characteristics of a given `transition' disc. However, no model or combination of models proposed can explain the observed population of `transition' discs. While photoevpoartion will ultimately destroy the disc, clearing it from inside-out, (naturally creating a `transition' disc phase), it cannot explain the full population of observed `transition' discs (Alexander \& Armitage 2009, Owen et al. 2011, 2012).

In fact, there is good evidence that observed `transition' discs do not represent a homogeneous population, but may contain several populations with different origins and lifetimes \citep[e.g.][]{merin10,OC12,espaillat14}. \citet{OC12} showed that there are two distinct populations of `transition' discs, with different properties that can be separated by their mm-flux. Owen \& Clarke (2012) demonstrated there is a population of `transition' discs at very-low mm fluxes (often the lowest mm fluxes of all protoplanetary discs); these discs have small ($<10$ AU) holes and low accretion rates ($<10^{-9}$ M$_\odot$ yr$^{-1}$). There is a second population of `transition' discs with high mm-fluxes (often with the highest mm-fluxes of all protoplanetary discs); these discs have large hole sizes ($>10 AU$) and high accretion rates ($10^{-9}-10^{-8}$ M$_\odot$ yr$^{-1}$). The characteristics of the `transition' discs with low mm-fluxes are consistent with the concept of a disc in transition from primordial to cleared - if one assumes that mm flux is a proxy disc mass, which declines with time - as these discs appear to be entering a `transition' disc phase at the end of the their lifetimes. Furthermore, comparing this low-mm flux population alone to the X-ray photoevaporation scenario indicates good agreement between these `transition' discs and the model predictions (Owen et al. 2011, 2012; Owen \& Clarke, 2012).

However, it is the population of `transition' discs at high mm-fluxes that still remains puzzling to understand. For photoevaporation to trigger disc clearing and create a transition disc, the accretion rate must drop below a threshold value (typically $\sim 10^{-9}$ M$_\odot$ yr$^{-1} $) and the hole always develops around $\sim 1~$AU (M$_*/{\rm M_\odot})$; meaning a `transition' disc with a hole at $\sim 20 AU$ and accretion rate of $10^{-8}$ M$_\odot$ yr$^{-1}$ is difficult to fit with a photoevaporative origin. Recently it has been suggested that a significant dead-zone \citep[e.g.][]{morishima12,bae13}, or an embedded planet \citep{rosotti13} may trigger photoevaporation at higher accretion rates and larger hole sizes. Although, it still remains difficult to reconcile the required dust depletion in the inner disc with these ideas. 

The original physical interpretation of a `transition' disc inferred from the  Spectral Energy Distributions (SEDs)  of an inner disc which is heavily dust depleted, but still contains a significant gas reservoir (to power the observed accretion), while sharply switching to a outer disc with a significant mass in gas and dust still remains today \citep[e.g.][]{calvet02,calvet05}. Many `transition' discs with large hole sizes have now been imaged in the sub-mm \citep[e.g][]{brown09,hughes09,andrews11,irs48} with measured holes sizes that agree with those inferred from the SED, indicating the interpretation of a dust depleted cavity is correct (at least for large dust particles with radii $s\sim 1$~mm). Detailed SED modelling \citep[e.g.][]{espaillat07,espaillat08,espaillat10} and MIR imaging \citep{geers07} have further confirmed that the inner disc\footnote{Some `transition' discs do present with a small optically thick inner disc at small radius $R\sim 0.1$ AU, termed `pre-transition' discs by \citet{espaillat07}, a distinction we are not concerned with here.} is also depleted of small dust particles, indicating the observed cavities must be dust-poor for particles with sizes $s\lesssim 1$~mm. Furthermore, the radial transition from optically thin inner cavity to optically thick outer disc still remains un-resolved, indicating the transition is sharp ($<10$~AU) and is inconsistent with the predictions of grain growth alone \citep{birnstiel12}. New {\it ALMA} observations of several `transition' discs have indicated that, as expected (from the measured accretion rates), the dust-poor cavities in `transition' discs contain a significant gas disc (Van der Marl et al. 2013; Perez et al. 2014) and in some cases the outer dust disc shows significant azimuthal asymmetries in the sub-mm images \citep{irs48,isella13}.

To explain the gas rich, dust-poor, inner regions of `transition' discs requires that the small dust particles are depleted by around $10^{-4}$ from standard primordial values \citep[e.g.][]{zhu11}. One of the most commonly invoked mechanisms is an embedded massive planetary companion, which creates a gap in the disc. Invoking a planetary companion is appealing since it naturally produces a leaky barrier to the gas \citep{calvet05,rice06,lubow06,dodson11,gressel13} allowing ongoing accretion, as well as driving azimuthal asymmetries in the outer disc, such as the Rossby wave instability \citep[e.g.][]{lovelace99,lin12,LL13}. Furthermore, the pressure bump caused by a gap-opening planet is strong enough to trap dust particles with non-dimensionless stopping times ($\tau_s$) of order unity \citep[e.g.][]{rice06,pinilla12,zhu12,zhu13}. Hole sizes of $\sim 20$ AU and accretion rates of $\sim 10^{-8}$ M$_\odot$ yr$^{-1}$ imply that particles with $\tau_s\sim 1$ are approximately $1~$mm in size, indicating that mm size particles will easily be trapped by a planetary gap \citep{rice06,zhu12} and can explain the observed mm images of `transition' discs (Pinilla et al. 2012). However, dust particles with smaller sizes (in particular those which dominate the NIR opacity $s<1~\mu$m) which have non-dimensional stopping times $\tau_s\sim10^{-3}-10^{-4}$ (indicating they will be tightly coupled to the gas) will not be trapped in the pressure bump created by the planet, instead following the gas through the gap and into the inner disc. Detailed two-fluid simulations by Zhu et al. (2012) demonstrated this explicitly and concluded while the planetary scenario could explain the observed hole radii, accretion rates and mm images, it would still have an optically thick inner disc (due to the small dust) and present with an SED consistent with a primordial disc. Thus, in-order to rescue the planetary scenario one must include an additional effect (other than pressure trapping) in order to remove the small dust particles from the inner disc and thus explain all the observed features of a `transition' disc, including the SED. Furthermore, models which invoke multiple planets to carve out a large gap \citep[e.g.][]{dodson11,zhu11} fail to explain all the characteristics. Zhu et al. (2011) showed that in the multiple planet scenario one could use planets to reduce the surface density in the inner disc, reproducing the observed SED and mm image, but failed to reproduce the observed accretion rates; modifying the model to match the observed accretion rates Zhu et al. (2011) found the model could no longer reproduce the NIR dip in the SED. 

A hint to the possible solution is that the observed accretion rates onto the star are high. This also implies a comparable accretion rate onto the embedded planet. In fact, this is backed-up by the exciting discovery of a low-mass accreting object inside the gap of the `transition' disc HD 142527 \citep{close14}, which exhibits a significant accretion luminosity. As we will discuss, this additional accretion luminosity (from the forming planet) can, in certain cases, be the dominant force on small dust particles and provide the missing mechanism to remove the small dust particles from the inner disc. This article is organised as follows: in Section~2, we lay the theoretical basis for the role radiation pressure from an accreting planet may play on the dynamics of small dust particles. In Section~3, we developed a coupled 1D secular gas and dust model and present the results of numerical calculations in Section 4. In Section~5, we use our results to compute synthetic observations, and discuss our results and model along with the caveats in Section~6, finally summarising in Section~7. This article also includes two appendices: Appendix~A discusses how to include planetary radiation pressure in a secular model and Appendix~B covers the tests of the numerical method.

\section{Overview}
One of the most difficult aspects to explain about `transition' discs with large ($>10$ AU) cavities is their observed accretion rate, which is comparable to primordial discs \citep[e.g.][]{espaillat14}. An appealing aspect of explaining `transition' discs with an embedded planet is that it naturally produces a leaky gap, which is known to trap mm sized dust particles \citep{rice06}. The accretion rate through the gap and into the inner disc is comparable to the accretion rate into the gap, and by construction comparable to the accretion rate onto the planet \citep[e.g.][]{lubow06,gressel13}. Following (Lubow \& D'Angelo, 2006) we define the mass-loss rate onto the planet ($\dot{M}_p$) in terms of the accretion rate into the inner disc ($\dot{M}_{\rm inner}$), and therefore onto the star ($\dot{M}_*$) as:
\begin{equation}
\dot{M}_{\rm p}=E\dot{M}_{\rm inner}
\end{equation}
Conservation of mass across the gap implies that the accretion rate onto the planet and into the inner disc, can be given in terms of the accretion rate into the gap from the outer disc ($\dot{M}_{\rm out}$) as:
\begin{eqnarray}
\dot{M}_p&=&\frac{E}{1+E}\dot{M}_{\rm out}\label{eqn:mdot_p}\\
\dot{M}_{\rm inner}&=&\frac{1}{1+E}\dot{M}_{\rm out}\label{eqn:mdot_in}
\end{eqnarray}
Since simulations suggest $E\sim 3-10$ (e.g. Lubow \& D'angleo, 2006) and given the observed accretion rates onto the star in `transition' discs are $\dot{M}_*\sim10^{-9}-10^{-8}$ M$_\odot$ yr$^{-1}$, then a planetary origin for `transition' discs would also suggest $\dot{M}_p\gtrsim 10^{-9}-10^{-8}$ M$_\odot$ yr$^{-1}$. In the following, we will make the assumption that `transition' discs with large cavities contain a massive planet ($M_p \gtrsim$ M$_J$) and that $E\gtrsim 1$ as the simulations suggest. 

Such a high accretion rate onto the planet will necessarily lead to a high accretion luminosity, which will be larger than the planet's intrinsic luminosity. For the scenario considered here this implies an accretion luminosity {\bl - estimated assuming the accreting material is free-falling onto the planet -}  of:
\begin{eqnarray}
L_p&=&\frac{GM_p\dot{M}_p}{R_p}=7\times10^{-3}{\rm L}_\odot\, E \left(\frac{M_p}{3\,{M_J}}\right)\nonumber \\ &\times &\left(\frac{\dot{M}_*}{10^{-8}\,{\rm M}_\odot\,{\rm yr}^{-1}}\right)\left(\frac{R_p}{10^{10}\, {\rm cm}}\right)^{-1}\label{eqn:p_lum}
\end{eqnarray} 
where $M_p$ is the planet and $R_p$ the planet radius. {\bl In deriving this expression we have made use of Equation~\ref{eqn:mdot_p} and implicitly assumed $\dot{M}_{\rm inner}=\dot{M}_*$ (which is true for a disc in steady-state)}. Thus, if we compare the bolometric flux received at the gap edge (taken to be a Hill radius - $R_H=a(M_p/3M_*)^{1/3}$ - from the planet, where $a$ is the separation and $M_*$ the stellar mass) compared to the star, {\bl assuming spherical dilution of the radiation} we find:
\begin{eqnarray}
\frac{F^p}{F^*}&=& 0.7 E \left(\frac{M_p}{3\,{M_J}}\right)^{1/3}\left(\frac{\dot{M}_*}{10^{-8}\,{\rm M}_\odot\,{\rm yr}^{-1}}\right)\left(\frac{R_p}{10^{10}\, {\rm cm}}\right)^{-1}\nonumber \\ &\times & \left(\frac{L_*}{L_\odot}\right)^{-1}\left(\frac{M_*}{M_\odot}\right)^{2/3}
\end{eqnarray} 
where $L_*$ is the star's bolometric luminosity. Therefore, at the gap edge we find that the accretion flux from the planet is comparable to, if not in excess of, the bolometric flux from the star. An obvious consequence of such a radiative flux is added feedback from radiation pressure. Since the dominant opacity source in protoplanetary discs is from dust particles then this radiation pressure will act on the dust particles. The flux-mean opacity for an individual spherical dust particle is given by:
\begin{equation}
\kappa=\frac{3Q}{4\rho_ds}
\end{equation}
where $\rho_d$ is the dust particle density, $s$ the dust particle radius and $Q$ is the radiative efficiency. For a perfect black-body dust grain $Q=1$ when $\lambda\ll s$ and $Q=(s/\lambda)^2$ when $\lambda \gg s$. We can estimate the magnitude of the radiation pressure due to a planet's accretion luminosity as $a^{\rm rad}=\kappa F^p/c$, where $c$ is the speed of light and compare it to the other sources of acceleration on a dust particle. Therefore, to determine the appropriate value of $Q$ we must estimate the radiation temperature of the accretion luminosity. In either the viscous boundary layer model, magentospheric accretion from a cirucumplanetary disc, or pure bondi accretion scenario we can roughly {\bl estimate the temperature of the radiation emerging from the accreting material at the planet's surface energetically as $3/2k_b T_{\rm acc}= GM_pm_H/R_p$ (where $k_b$ is the Boltzmann constant, and $m_H$ is the mass of a hydrogen atom), which yields $T_{\rm acc}\sim 3\times10^{5}$~K for the planets considered here. Wien's displacement law then gives an estimate of the photon wavelength of $\lambda \sim 0.01~\mu$m. Ultimately this radiation maybe reprocessed to longer wavelengths by extinction material; however, for the sake of simplicity for the initial calculations presented here we assume the wavelength of the radiation is shorter than the size of the particles of interest.
}
 Thus, for all the dust particles we set $Q$ to unity with respect to the planetary accretion flux.

We now want to compare the magnitude of radiation pressure on a small (s$\sim$ 0.1 $\mu$m, $\rho_d=2$ g cm$^{-3}$) dust particle orbiting at the gap edge to the other forces that govern the dynamics of a dust particle. Evaluating the radiation pressure pressure at the gap edge we find a radiative acceleration of:
\begin{eqnarray}
a^{\rm rad}&=&3\times10^{-3}{\, \rm cm\,\,s^{-1}}E\left(\frac{s}{0.1{\rm \, \mu m}}\right)^{-1}\left(\frac{M_p}{3{\,\rm M_J}}\right)^{1/3} \left(\frac{M_*}{\rm M_\odot}\right)^{2/3}\nonumber \\
&\times &\left(\frac{\dot{M_*}}{10^{-8}{\,\rm M_\odot \,yr^{-1}}}\right) \left(\frac{R_p}{10^{10}{\,\rm cm}}\right)^{-1}\left(\frac{a}{20{\rm \, AU}}\right)^{-2}
\end{eqnarray}
comparing this to the radiation pressure from the star:
\begin{eqnarray}
a^{\rm rad}_*&=&8\times10^{-5}\,{\rm cm\,s}^{-2} \left(\frac{s}{0.1\,\mu{\rm m}}\right)\left(\frac{T_*}{4000{\rm \,K}}\right)^2\nonumber \\ &\times &\left(\frac{L_*}{{\rm L_\odot}}\right)\left(\frac{a}{20{\, \rm AU}}\right)^{-2}
\end{eqnarray}
where since the stellar black-body peaks at $\lambda > s$ we have used the perfect spherical blackbody approximation to suitably scale the radiative efficiency. The instantaneous radiative acceleration from the planet is large, thus, in order to asses when it might be dominant we must compare it to the the other main `acceleration' acting on the dust particle; namely, due to drag from gas advection.  

Adopting the Epstein regime, the acceleration on the dust particle due to dust advection is approximately:
\begin{equation}
a^{\rm adv}=\frac{\dot{M_*}c_s}{4\pi^2 (H/R)a^2\rho_ds}\label{eqn:drag}
\end{equation}
where $c_s$ is the local sound speed, and $H$ is the disc scale height at a radius $R=a$. Additionally, noticing that both the acceleration due to dust advection and radiation pressure scale identically with dust density, accretion rate, and particle size we can place a constraint on the value of $E$, such that the radiation pressure is larger than the acceleration due to dust advection.
\begin{eqnarray}
E &\gtrsim& \frac{1}{2}\left(\frac{H/R}{0.1}\right)^{-1}\left(\frac{a}{20\,{\rm AU}}\right)^{-1/4}\left(\frac{M_p}{3\,{\rm M_J}}\right)^{1/3}\nonumber \\ &\times & \left(\frac{R_p}{10^{10}\,{\rm cm}}\right)^{-1}\left(\frac{M_*}{\rm M_\odot}\right)^{2/3}\label{eqn:ratio}
\end{eqnarray}
{\bl where since the disc is passivly heated \citep[e.g.][]{chiang97,dalessio01}, we adopt a  $T\propto R^{-1/2}$ temperature profile \citep[e.g.][]{kenyon87}}. Thus, we see for the values of $E$ typically found from simulations the instantaneous radiation pressure from an accreting planet can dominate over the advection of small dust particles by a large factor. We strongly caution that this analysis is a very rough guide and ignores two important (and in practical considerations dominant) additional considerations. Firstly, that the instantaneous radiation pressure only acts over part of the dust particles orbit, while the dust drag due advection acts of the entire orbit, necessarily weakening the role of radiation pressure. Secondly the planetary gap will modify the surface density profile in the gap away from the simple steady-state ($\dot{M_*}=3\pi\nu\Sigma$) form we have used to estimate the advective drag force in Equation~\ref{eqn:drag}. This tends to weaken the strength of the advective drag force by reducing the coupling between the dust and gas through a reduction in the gas density close to the planetary gap (a very important consideration we will discuss in detail in Section~6). Ultimately, the competition between these two additional considerations determines whether the radiation pressure can stop the accretion of the small dust particles. This issue is the main aim of this work and we will answer this question through numerical simulations presented in Section 3. Finally, we note that for larger particle sizes, with non-dimensional stopping times $\tau_s\sim 1$, that dust drag due to the differential azimuthal velocity between the gas and dust becomes dominate over dust-drag due to gas advection and radiation pressure. Thus, we expect that for dust particles with $\tau_s\sim 1$ that their dynamics will be governed by the pressure distribution of the gas disc.

Given the strength of the radiation pressure, this gives hope that for  sufficiently high accretion rates, radiation pressure may be able to hold back the dust particles, and warrants further study in this work. Since the small dust particles are tightly coupled they quickly transfer their excess momentum to the gas. Thus, one needs to check whether this will have a dynamical consequence on the gas itself. Adopting a dust particle number density distribution of the form $n(s){\rm d}s\propto s^{-p}{\rm d}s$, the radiative acceleration on the gas will be given by:
\begin{equation}
a^{\rm rad}_{\rm gas}\approx
\begin{cases}
Xa^{\rm rad}_{\rm dust}(s=0.1\mu{\rm m})\left(\frac{s_{\rm max}}{0.1{\rm \, \mu m}}\right)^{-1} & \!\!\text{$p<3$}\\
Xa^{\rm rad}_{\rm dust}(s=0.1\mu{\rm m})\left(\frac{s_{\rm max}}{0.1{\rm \, \mu m}}\right)^{-1}\left(\frac{s_{\rm min}}{s_{\rm max}}\right) & \!\!\text{$3<p<4$}\\
Xa^{\rm rad}_{\rm dust}(s=0.1\mu{\rm m})\left(\frac{s_{\rm min}}{0.1{\rm \, \mu m}}\right)^{-1} &\!\! \text{$p<4$}
\end{cases}
\end{equation}
where $X$ is the dust-to-gas mass ratio and $s_{\rm min}$ \& $s_{\rm max}$ are the minimum and maximum dust particle size respectively. Therefore, we see for any dust distribution with $p>4$, (namely the mass in the dust distribution is dominated at large particle sizes) radiation pressure will not inject significant momentum into the gas, and to first order the radiation pressure from the planet will only affect the dynamics of the dust distribution and not the gas, provided the dust distribution has grown to sizes $s_{\rm max}\gg 0.1\,\mu$m. Given protoplanetary discs are expected to have $s_{\rm max}\gtrsim 1$ mm and $p>4$ \citep[e.g.][]{birnstiel10,pinilla14} we can at this stage safely ignore the effect of the radiation pressure on the gas. 

{\bl \subsection{Consequences of vertical structure}}

Having seen that radiation pressure from an accreting planet may have a dynamical consequence on dust particles close to the gap edge, a concern is whether any circumplanetary disc may shield the outer regions of the protoplanetary disc from direct lines of sight to the planet. This can be considered by comparing the scale heights of any circumplanetary disc ($H_{\rm cd}$) to the scale height of the protoplanetary disc ($H$). The protoplanetary disc will be passively heated by the central star at large radius {\bl \citep{chiang97,dalessio01} so the scale height of the protoplanetary disc is given by \citep[e.g.][]{kenyon87}}:
\begin{equation}
\frac{H}{R}=0.04\left(\frac{R}{1{\rm \, AU}}\right)^{1/4}
\end{equation}
In contrast, the circumplanetary disc is an active disc, heated by accretion, with a scale height of $H_{\rm cd}/R\approx0.3$, which is relatively insensitive to the determining parameters in the range of interest \citep{martin11}. Furthermore, the circumplanetary disc is truncated near the orbit crossing radius of 0.4$R_H$ \citep{martin11}. Thus, the height of the circumplanetary disc at its outer edge is:
\begin{equation}
H_{\rm cd}=0.2 {\rm \, AU}\left(\frac{a}{20{\rm \, AU}}\right)\left(\frac{M_p}{3{\rm \, M}_{\rm J}}\right)^{1/3}
\end{equation}
comparing this to the scale height of the protoplanetary disc at the orbit of the planet:
\begin{equation}  
H=1.7{\rm \, AU}\left(\frac{a}{20{\rm \, AU}}\right)^{5/4}
\end{equation}
we see the circumplanetary disc is unable to shade the protoplanetary disc from  direct lines of sight with the planet. Thus, the photons produced by accretion are able to directly impinge upon the edge of the protoplanetary disc. 

{\bl In addition, it is well known dust-particles can sediment towards the mid-plane of the protoplanetary disc while conversely being lofted by turbulence, thus it is important to check that the small grains will remain well mixed vertically in the disc. At a height $z\sim H\ll a$ the vertical components of gravity and radiation pressure are given by:
\begin{eqnarray}
F^{g}_z&=&-\frac{GM_*m_d}{a^3}z\\
F^{\rm rad}_z&=&\frac{m_d\kappa L_p}{4\pi c\left(R_H^2+z^2\right)^{3/2}}z
\end{eqnarray}
comparing gravity and radiation pressure to drag force vertically and assuming the dust particles are tightly coupled to the gas (hence adopting the terminal velocity approximation) then we find a settling time-scale of:
\begin{equation}
t_{\rm set}=\frac{\exp\left(-z^2/2H^2\right)}{\Omega\tau_s(1-\beta)}
\end{equation}
where $\beta$ represents the fractional reduction in vertical gravity due to radiation pressure given by:
\begin{equation}
\beta=\frac{\kappa L_p}{4\pi GcM_*\left[\left(M_p/3M_*\right)^{2/3}+\left(z/a\right)^2\right]^{3/2}}
\end{equation} 
 Comparing this settling time to the turbulent lofting time-scale $t_{\rm loft}\approx z^2/\alpha c_s H$, we can estimate the height $z_{\rm dust}$ of the dust layer in the disc by equating $t_{\rm set}$ and $t_{\rm loft}$ as:
\begin{equation}
\frac{\alpha}{2\tau_s(1-\beta)}=\left(\frac{z_{\rm dust}^2}{2H^2}\right)\exp\left(\frac{z_{\rm dust}^2}{2H^2}\right)
\end{equation}
which can be expressed in closed-form as:
{\rd \begin{equation}
z_{\rm dust}=H \sqrt{2W_0\left(\frac{1}{2S(1-\beta)}\right)}
\end{equation}}
where $S$ is the ratio of the viscosity to dimensionless stopping time {\rd ($\tau_s/\alpha$ e.g. \citealt{jacquet12})} and $W_0$ is the Lambert W function\footnote{$W_0(x)$ is logarthimically divergent at large $x$.}. For the small particles we are interested in here {\rd $S\ll 1$}, $z_{\rm dust} > H$. Thus, even without the added help of radiation pressure, the small dust is well mixed to many scale heights and is not settled into the mid-plane. Therefore, the photons produced by accretion are able to directly impinge upon the small dust particles at the edge of the disc. We note in parsing that at large heights the small dust particles are no longer tightly coupled to the gas. Under certain circumstances, radiation pressure from the planet may be able to drive a `dust-particle wind' from several scale heights. Such a `wind' will have obvious implications for scattered light observations of `transition' discs such as those resulting from the SEEDs project \citep[e.g.][]{dong12}.
}

In this section we have set the ground work and seen that the radiation pressure from an accreting planet may have a dynamical consequence on the small dust particles at the outer edge of the planetary gap. In order to assess whether, for any sensible scenario, this can resolve the `transition' disc conundrum we must turn to numerical calculations. 

\section{One-dimensional numerical models}

In order to model the `transition' disc problem we must build a numerical model which allows us to asses whether the addition of planetary accretion luminosity can help explain the observed features. We choose to model the problem in a simple way, adopting a 1D radial model as has been used in previous studies of gas and dust discs with special regard to `transition' discs \cite[e.g.][]{alexander07,alexander09,zhu11,owen11,AP12,birnstiel12,pinilla12}. 

\subsection{Secular models}
The governing equations for the gas and dust discs with an embedded planet are given by \cite[e.g.][]{clarke88,takeuchi02,lodato04,alexander07,alexander09,owen11}:
\begin{equation}
\frac{\partial \Sigma_g}{\partial t}=\frac{3}{R}\frac{\partial}{\partial R}\left[R^{1/2}\frac{\partial}{\partial R}\left(R^{1/2}\nu\Sigma_g\right)-\frac{2\Lambda\Sigma_gR^{3/2}}{\sqrt{GM_*}}\right]\label{eqn:gas}
\end{equation}
and
\begin{equation}
\frac{\partial \Sigma_d^i}{\partial t}=-\frac{1}{R}\frac{\partial}{\partial R}\left[R\Sigma_d^iv_d^i-\frac{\nu}{Pr}{\rd R}\Sigma_g\frac{\partial}{\partial R}\left(\frac{\Sigma_d^i}{\Sigma_g}\right)\right]\label{eqn:dust}
\end{equation}
where the super-script $i$ refers to a dust particle of size $s^i$, $\nu$ is the turbulent viscosity and $Pr$ is the Prandtl number describing the ratio of the turbulent viscosity to the dust diffusion due to the turbulence. $\Lambda$ is the torque resulting from the planet, where we use the symmetric form used by \citet{trilling98,armitage02,lodato04,alexander09,AP12}; for a planet to star mass ratio of $q$, $\Lambda$ is given by:
\begin{equation}
\Lambda=
\begin{cases}
-\frac{q^2GM_*}{2R}\left(\frac{R}{\max(H,|R-a|)}\right)^4 & \text{if $R<a$} \\
\frac{q^2GM_*}{2R}\left(\frac{a}{\max(H,|R-a|)}\right)^4 & \text{if $R>a$} 
\end{cases}
\end{equation}

Since protoplanetary discs are primarily passively heated (e.g. Chiang \& Goldreich, 1997, D'alessio et al. 2001), a constant `$\alpha$' viscosity requires $\nu\propto R$ (e.g. Hartmann et al. 1998, Alexander et al. 2006, Owen et al. 2011). Therefore, we set $\nu=\nu_0R$ where $\nu_0=\alpha(H/R)^2\Omega R^2$ in all our simulations. In all calculations we set $\alpha=0.0065$; $H/R$ is normalised to 0.04 at 1 AU and we choose to set the Prandtl number to unity (e.g. Clarke \& Pringle 1988; Alexander \& Armitage 2007). The radiation pressure on the dust is included through its effect on the dust velocity via Equation~\ref{eqn:d_vel}. 

In order to determine $v_d^i$, one needs to account for the additional impact on the dust due to the radiation pressure. The equations governing the evolution of a dust particle in a gas disc including radiation pressure\footnote{Note here we neglect the contribution due to the planetary torque on the dust particles; for tightly coupled particles it is easy to show that the effect of the torque is of order $\tau_s^2$, where as the radiation pressure and dust drag are of order $\tau_s$. Thus the torque on the dust is negligible for tightly coupled particles, but certainly not for particles with $\tau_s>1$ - see Zhu et al. 2013.} are: 
\begin{equation}
\frac{dv_R}{dt}=\frac{v_\phi^2}{R}-\Omega^2R-\frac{1}{t_s}\left(v_R-u_R\right)+a_R^{\rm rad}\label{eqn:dustR}
\end{equation}
and 
\begin{equation}
\frac{dRv_\phi}{dt}=-\frac{R}{t_s}\left(v_\phi-u_\phi\right)+Ra_\phi^{\rm rad}\label{eqn:dustphi}
\end{equation}
where $t_s$ is the stopping time given by $t_s=\tau_s/\Omega$. Since we are primarily interested in tightly coupled particles $\tau_s<1$, then the net effect of the planetary radiation pressure will be to impart an impulse on the dust particle every orbit, knocking it onto a slightly different orbit. Therefore, we adopt a 1D orbit averaged (we are interested in the secular evolution) approach and can average the impulse imparted by the radiation pressure from the planet over an entire orbit, and use Equations \ref{eqn:dustR} \& \ref{eqn:dustphi} to calculate the secular evolution of the dust particles. We do this by replacing $a^{\rm rad}$ by its orbit averaged value $\left<a^{\rm rad}\right>$ in Equations \ref{eqn:dustR} \& \ref{eqn:dustphi}; noticing by symmetry $\left<a^{\rm rad}_\phi\right>=0$ we can then proceed and solve for $v_R$.  Following standard derivations \citep[e.g][]{takeuchi02,armitagebook} we can include the radiation pressure term. One takes the dust particles to move through a succession of circular orbits, so we may simplify the azimuthal equation to:
\begin{equation}
v_\phi-u_\phi=-\frac{1}{2}t_sv_R\Omega
\end{equation}
Thus, solving for the radial dust velocity we find:
\begin{equation}
v_R=\frac{u_R\tau_s^{-1}-\eta \Omega R +\langle a^{\rm rad}_R\rangle/\Omega}{\tau_s+\tau_s^{-1}}\label{eqn:d_vel}
\end{equation}
where $\eta$ is a measure of the gas pressure gradient given by:
\begin{equation}
\eta=-\frac{{\rm d}\log P}{{\rm d}\log R}\left(\frac{c_s}{v_K}\right)^2
\end{equation}
We discuss in Appendix~A how to perform the orbit averaging of the planetary radiation pressure, we evaluate $\left<a^{\rm rad}_R\right>$ and also present a closed form solution in the optically thin limit.

\subsection{Numerical method}

Operationally, Equations \ref{eqn:gas} \& \ref{eqn:dust} are integrated  explicitly using a scheme that is second order in space and first order in time; the flux terms are reconstructed using a van-Leer limiter. Furthermore, to make the time-step numerically manageable we follow Lodato \& Clarke (2004) and Alexander \& Armitage (2009) and smooth both the planetary torque and the radiation pressure inside the planet's hill sphere and we do not attempt to model the disc properties in this region  (essentially in the regions where the flow is no-longer 1D). In order to model the flow across the planetary gap, and onto the planet, we adopt a `leakage' prescription similar to those previously used in the literature \citep[e.g.][]{alexander09,alexander12,AP12}. Since the `leakage' is applied outside the planet's hill sphere and the smoothing takes place inside, the nature of this smoothing does not affect the calculations \citep{alexander09}.
The `leakage' is included as a sink and source term outside and inside the planet's hill sphere that moves dust and gas from one-side of the planet's orbit to the other. In order to determine the gas leakage rate we measure the steady-state viscous rate ($\dot{M}_{\rm out}$) at $3a$ \citep{alexander09}. We then assume that the mass-flux across the gap and into the inner disc is given by Equation~\ref{eqn:mdot_in} and the accretion rate onto the planet is given by Equation~\ref{eqn:mdot_p}. For the dust we weight the gas leakage rate ($\dot{M}_{\rm out}$) by the dust concentration at the gap edge, and use the same value of $E$ to decide how much dust is accreted by the planet. It is also necessary to apply a time-step limit to the `leakage' source and sink terms so that they do not account for a $>1\%$ change in the surface density.
 To determine the orbit averaged radiation pressure on the dust we proceed as follows: at the beginning of each time-step we determine the accretion rate onto the planet and calculate the accretion luminosity using Equation~\ref{eqn:p_lum}. For each grid cell we determine the optical depth on each cell face by integrating the radiative transfer equation as (assuming the dust is vertically well mixed):
\begin{equation}
\tau^i_{\rm face}=\sum_{j=k_p+1}^{i-1}\frac{\Sigma_g^j}{\sqrt{2\pi}H_j}\sigma^{j-1}\Delta R^{j-1}
\end{equation}
where $k_p$ is the position of the planet, $\sigma^{j}$ is the cell centred cross-section of the dust distribution and $\Delta R^j$ is the radial cell size.  We calculate $\sigma^{j}$ independently for each cell assuming each dust particle to have a geometric cross-section, the local dust size distribution and the local dust-to-gas mass ratio. We then evaluate the orbit averaging integral (Equation~A2) numerically in order to determine the orbit averaged radiation pressure.

 Furthermore, in order to include sufficient resolution around the planet we make use of static mesh refinement, where we increase the number of grid cells, by splitting each cell of the `mother' grid into 10 extra cells within 3 hill radii of the planet. The `mother' grid is logarithmically spaced between 0.02 and 200 AU and composed of 400 cells, this results in approximately 40-50 cells per hill radii in the vicinity of the planet. Since this high resolution requirement makes an evolutionary calculation computationally infeasible, we model a `quasi-steady-state' version of the problem. We do this by neglecting the planet's migration and evolving the model until a steady state is achieved in the gas distribution. This is done by using inflow boundary conditions at the outer boundary, where gas and dust is injected into the grid at a constant rate. At the inner boundary we apply a zero torque boundary condition and set $\Sigma=0$. In all cases the gas initially has a $\Sigma\propto R^{-1}$ profile (appropriately scaled for the incoming accretion at the outer boundary) and no dust is present. We then evolve the gas only for 1~Myr with the planet to obtain steady-state in the gas. After 1~Myr we introduce dust at the outer boundary, allowing the simulation to evolve for a further 1~Myr. 

While steady-state is always achieved for the gas, this will not be the case for trapped dust particles, where a `quasi-steady-state' is achieved after 1~Myr. The concentration of dust particles in the inner disc slowly increases as the concentration of dust trapped at the outer edge grows, increasing the concentration gradient at the gap. This is a slow diffusive process and all our measurements are made at 2~Myr. Allowing the simulation to run for several 100~Myrs,  a steady-state (where the rate dust particles entering the grid at the outer boundary equals the rate exiting at the inner boundary - excluding the planet's accretion) can be reached; however, such a situation results in an unphysically large dust concentration ($\sim 100-1000$) at the gap edges. The tests we used to ensure the numerical scheme is behaving accurately are discussed in Appendix~\ref{App:code_test}.

In all models we adopt a stellar mass of $1\,{\rm M}_\odot$, the dust distribution injected at the outer boundary has a dust-to-gas mass ratio of 0.01, with an MRN power-law distribution ($p=3.5$, \citealt{MRN}),  a minimum particle size of $s_{\rm min}=0.005\,\mu$m and a maximum particle size of $s_{\rm max}=1\,$mm. {\bl Such a choice is a reasonable starting point and is often the grain size chosen for radiative transfer modelling \citep[e.g][]{dalessio01,koepferl13}. It is well known that the dust distribution can evolve away from the MRN distribution \citep{birnstiel10,birnstiel12}. As such if the particle distribution becomes strongly peaked towards small particle sizes it will weaken the effect of radiation pressure; however if the particle distribution becomes strongly dominated by large particle sizes it will strengthen the role of radiation pressure.}
The dust distribution is followed using 25 size bins, logarithmically spaced between $s_{\rm min}$ and $s_{\rm max}$. {\bl We note that we do not include dust growth or fragmentation (a caveat we discuss in Section~6) and the dust distribution evolves only due to radial motion/trapping of individual dust species; therefore, our 25 size bins evolve independently.} Finally we pick the median value of $E$ from Lubow \& D'angleo (2006) of $E=6$ which is comparable with value used in similar 1D models of the planet disc interaction (Alexander \& Armitage 2009, Alexander \& Pascucci 2012).

\section{Results}
We calculate several models that have parameters comparable to observed `transition' discs, but unfortunately the high numerical overhead\footnote{This is primarily due to the short-time step required by the models, necessitated by the high resolution in the vicinity of the planet required to resolve the optical depth and concentration gradients.} restricts this parameter range to a small sub-set of the parameter space. The simulated parameters are shown in Table~\ref{tab:sim}. Taking  $\dot{M}_*=10^{-8}$ M$_\odot$ yr$^{-1}$, we simulate the evolution of gas and dust in a protoplanetary disc with a planet. Placing an embedded planet on a fixed circular orbit at 20 AU with various masses we follow the evolution of the gas and dust. For our `standard' model we consider a planet with a mass of 4.0 M$_J$, a radius of $10^{10}$ cm and do not include the radiation pressure feedback (simulation A). 
\begin{table*}
\centering
\begin{tabular}{l | c c c c c c}

\hline
Simulation & M$_p$ [M$_J$] & R$_p$ [cm] & $\langle a_R^{\rm rad}\rangle$ (Y/N) & $\dot{M}$ [M$_\odot$ yr$^{-1}$] & a [AU] & $E$ \\
\hline
\hline
A & 4.0 & $10^{10}$ & N & $10^{-8}$ & 20 & 6 \\
B & 4.0 & $10^{10}$ & Y & $10^{-8}$ & 20 & 6 \\
\hline
C & 2.0 & $10^{10}$ & Y & $10^{-8}$ & 20 & 6\\
D & 2.5 & $10^{10}$ & Y & $10^{-8}$ & 20 & 6\\
E & 3.0 & $10^{10}$ & Y & $10^{-8}$ & 20 & 6\\
F & 3.5 & $10^{10}$ & Y & $10^{-8}$ & 20 & 6\\
G & 4.5 & $10^{10}$ & Y & $10^{-8}$ & 20 & 6\\
\hline
H & 4.0 & $10^{10}$ & Y & $10^{-8}$ & 20 & 0.5 \\
\hline
\end{tabular}
\caption{\textup{Simulation parameters; {\bl $\langle a_R^{\rm rad}\rangle$ (Y/N), refers to whether radiation pressure from the planet was included in the simulation.}}}\label{tab:sim}
\end{table*}
\begin{figure}
\centering
\includegraphics[width=\columnwidth]{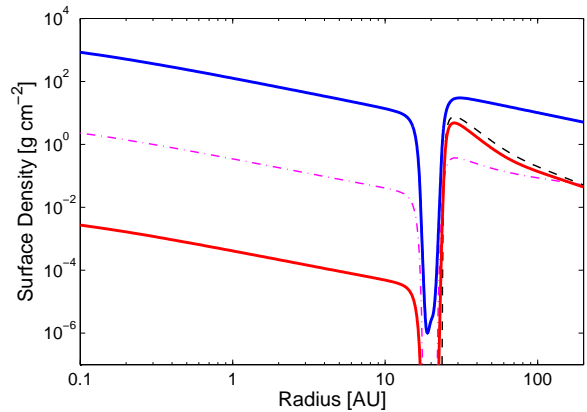}
\caption{The dust and gas distribution for simulation A: a disc with a 4.0M$_{\rm J}$ planet on a circular orbit at 20AU, without including the effect of radiative feedback on the dust. The thick blue line shows the gas surface density, while the thick red line shows the total dust surface density. The dot-dashed and dashed line shows the surface density of sub-micron ($\sim$0.1$\mu$m) and mm-sized ($\sim 1$ mm) dust particles respectively, {\bl note these are not plotted with physical units (e.g. g cm$^{-2}$), but rather; both dashed and dot-dashed lines are scaled such that their individual dust-to-gas ratios are 0.01 at large radius.} }\label{fig:no_arad}
\end{figure}

In Figure~\ref{fig:no_arad} we show the results of simulation A. This shows that the pressure gradient induced by the planet can trap large mm-sized dust particles, in agreement with previous studies \citep[e.g.][]{rice06,zhu12,zhu13,pinilla12}. However, the pressure gradient is unable to trap small sub-micron dust particles which freely flow across the gap and into the inner disc, essentially tracing the gas distribution. This simulation shows similar results to those presented by \citet{zhu12} using multi-dimensional two-fluid simulations. Essentially the time to advect small dust particles into and across the gap is much shorter than the time-scale on which they will feel the presence of the pressure bump and migrate towards it.

\begin{figure}
\centering
\includegraphics[width=\columnwidth]{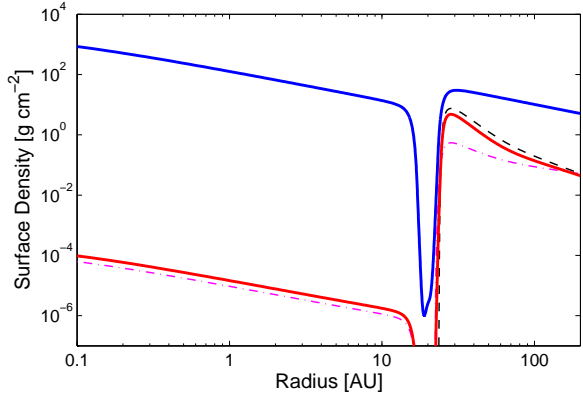}
\caption{The dust and gas distribution for simulation B: a disc with a 4.0 M$_{\rm J}$ planet on a circular orbit at 20AU, where the effect of radiative feedback from the accreting planet has been included. The thick blue line shows the gas surface density, while the thick red line shows the total dust surface density. The dot-dashed and dashed lines shows the density of sub-micron ($\sim$0.1$\mu$m) and mm-sized ($\sim 1$mm) dust particles respectively. Folowing Figure \ref{fig:no_arad} both lines are scaled such that their individual dust-to-gas ratios are 0.01 at large radius.}\label{fig:with_arad}
\end{figure}

In simulation B, we repeat simulation A ($M_p=4.0$ M$_J$), but this time include the effect of radiation pressure. The resulting gas and dust surface densities are shown in Figure~\ref{fig:with_arad}; here we see that, as before, the mm particles are trapped in the outer disc to a similar level to that found in simulation A (without radiation pressure). This means, as in simulation A, the dynamics of the mm sized particles are dominated by the pressure bump outside the planet, and the inclusion of radiation pressure has little effect on their dynamics. However, Figure~\ref{fig:with_arad} shows that the small sub-micron sized dust particles are significantly suppressed in the inner disc (we find a suppression factor of $< 10^{-4}$). Here the radiation pressure traps the dust particles at the outer edge of the gap, and the resulting concentration in the inner disc is set by a balance of advection, diffusion and radiation pressure at the gap edge.

\begin{figure}
\centering
\includegraphics[width=\columnwidth]{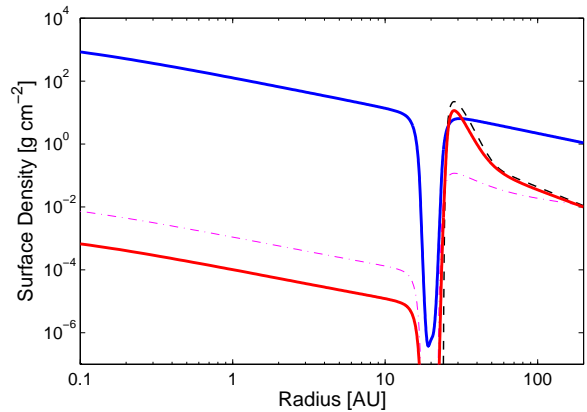}
\caption{Same as Figure~\ref{fig:with_arad} but for $E=0.5$.}\label{fig:lowE}
\end{figure}

Furthermore, to asses whether the planet needs to accrete a significant fraction of the gas entering the gap we repeat simulation B, but with $E=0.5$ in simulation H. The resulting surface density profiles are shown in Figure~\ref{fig:lowE}. With this lower value of $E$ we find that, as in the case with no radiation pressure, the mm-sized dust particles are trapped by the planetary pressure bump and the sub-micron grains still make it across the gap. Comparing simulations A,B \& H we find that while the sub-micron grains are suppressed in the inner disc for $E=0.5$ compared to the model with no radiation pressure (simulation A), the level of trapping for the sub-mm grains is not enough for the disc to give rise to a NIR dip in the SED and would still be classified as a primordial disc. Thus, we require the planet to be accreting the majority ($E\gtrsim 1$) of the gas entering the gap for the disc to appear as a `transition' disc.

\subsection{Variations in planetary mass}

\begin{figure}[t]
\centering
\includegraphics[width=\columnwidth]{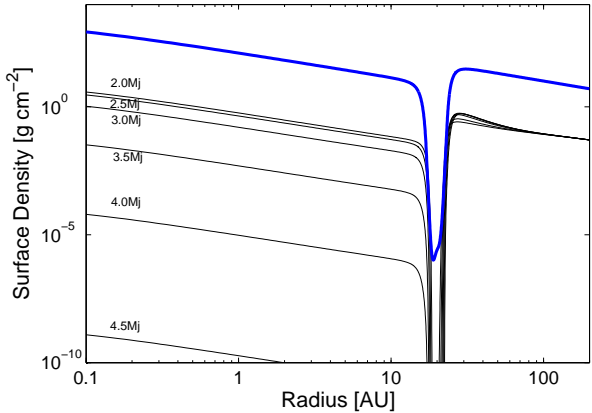}
\caption{Gas (thick line, shown only for $M_p=4.0$~M$_{\rm J}$ for reference) and sub-micron dust distribution (thin lines) shown for planets with masses in the range 2.0 to 4.5 M$_{\rm J}$ on circular orbits at 20 AU. As in previous Figures, the sub-micron ($\sim$0.1$\mu$m) dust particles' surface densities have been scaled such that their individual dust-to-gas ratios are 0.01 at large radius.}
\label{fig:planet_mass}
\end{figure}

Finally, to conclude the initial study of the simulations presented in this work, we investigate the role of the planet's mass (simulations B-G). We consider a range of planets between $2.0$ and 4.5 M$_{\rm J}$  all with radii of $10^{10}$ cm, on circular orbits at 20 AU. The resulting surface density profiles for the sub-micron dust are shown in Figure~\ref{fig:planet_mass} for Simulations B-G.

These simulations clearly show for planet's with masses above $\sim4$ M$_{\rm J}$, the dynamics of small sub-micron sized dust particles are governed by radiation pressure from the accreting planet and the inner disc becomes optically thin at NIR wavelengths, ultimately giving rise to the signature of a transition disc in the SED for an accretion rate of $\dot{M}=10^{-8}$ M$_\odot$ yr$^{-1}$. As in the comparisons between simulations A \& B we find that in all cases the dynamics of the larger mm-sized particles are governed by the properties of the pressure bump and large planets trap more mm-size grains; however, even planets with masses $\sim 2$ M$_J$ are able to trap enough mm grains to create a cavity in the mm images, as we will see in Section~6. While the radiation pressure at the gap edge does increase with planet mass (Equation \ref{eqn:p_lum}) this is not the major reason why the planet mass has a strong effect on the level of small dust particles entering the inner disc. An important fact that we could not include in our simple discussion in Section~2 is that bigger planets carve deeper gaps. This has two effects: i) the optical depth at the gap edge where the dust is trapped drops, and ii) the reduction in surface density reduces the dust-gas coupling making it easier for the radiation pressure to overcome the gas advection. These processes are discussed further in Section~6. 

\section{Observational Diagnostics}
In order to compare the simulations with current observations we calculate both the disc's spectral energy distribution and mm image. Since our simulations give us the dust distribution as a function of radius we can use it to calculate the opacity to the incoming stellar and re-radiated thermal emissions. Unlike the calculation of the radiation pressure term we cannot simplify the opacity of the dust particles and thus we use  tabulated values from \cite{laor93}, assuming a 50/50 mixture of graphite and silicate grains. 

\subsection{SED calculation}
In all cases we assume the disc is observed face-on and vertically isothermal. Since the goal of this section is to calculate representative SEDs, rather than accurate models for SED fitting, we do not use a full numerical radiative transfer approach, but rather estimate the disc's radial temperature profile and hence brightness analytically. We calculate a disc temperature assuming both an active disc (heated only by accretion - \citealt{shakura73}) and passive disc (heated only by stellar irradiation - \citealt{chiang97}), and choose the maximum of the two temperatures to be the disc's temperature. We also apply a minimum disc temperature of 10K and a dust sublimation temperature of 1500K. 
The passive temperature profile is taken to be:
\begin{eqnarray}
T(R)&=&290\,{\rm K}\left(\frac{R}{1{\,\rm AU}}\right)^{-1/2}\exp(-\tau_R^*)\nonumber \\ &+&200\,{\rm K}\left(\frac{R}{1\,{\rm AU}}\right)^{-1/2}\left[1-\exp(-\tau_R^*)\right]
\end{eqnarray}
where $\tau_R^*$ is the radial optical depth to stellar radiation used to smoothly move between the radially optically thin and thick limits. 
Thus, the luminosity of a face-on thin disc is calculated as:
\begin{equation}
L_\nu=\pi\int_0^{\infty}{\rm d}R\, 2\pi RB_{\nu}(T(R))\left[1-\exp(-\tau_\nu)\right]
\end{equation}
where $\tau_\nu$ is the vertical optical depth at frequency $\nu$ as a function of radius, and $B_\nu$ is the planck function.
\subsection{Millimetre images}
Most of the mm images to date have been imaged by the {\it Sub-Millimetre Array} at 880$\mu$m (e.g. Andrews et al. 2011), therefore we calculate our images at 880$\mu$m. We use the same radial temperature distribution used to calculate the SEDs and use our dust distribution to calculate the opacity at 880$\mu$m, and associated vertical optical depth. The source function for a face-on, thin disc is then approximately:
\begin{equation}
S_\nu=B_\nu\left[1-\exp(-\tau_{880})\right]
\end{equation}
In order to calculate the image we assume the dust distribution to be axis-symmetric at all particle sizes. Furthermore, since the gap features occur on smaller scales than typical observational resolutions, we further degrade the image by convolving it with an axis-symmetric Gaussian beam with a standard-deviation of 5~AU. 

\begin{figure}[t]
\centering
\includegraphics[width=\columnwidth]{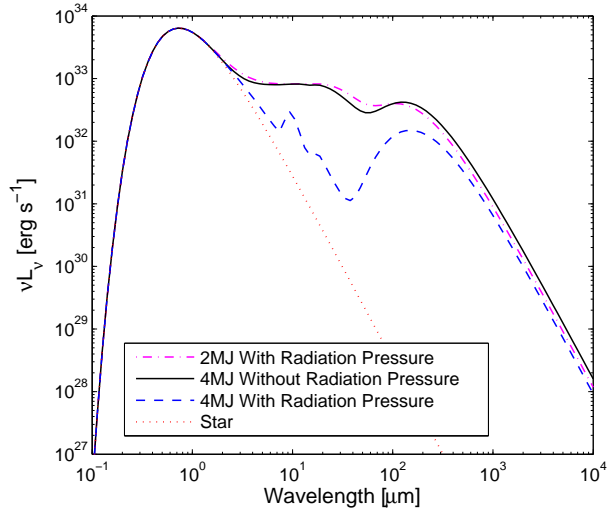}
\caption{Simulated spectral energy distributions of simulation A (solid), B (dashed), \& C (dot-dashed). The stellar spectrum is shown as the dotted line.}\label{fig:fig3}
\end{figure}

\begin{figure*}
\centering
\includegraphics[width=\textwidth]{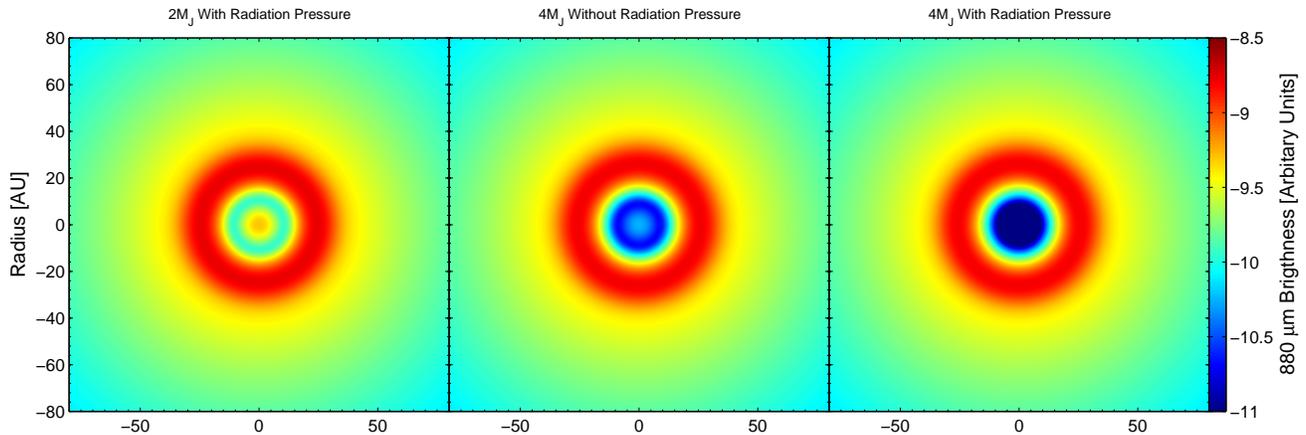}
\caption{Simulated 880~$\mu$m observations of simulations A (centre), B (right) \& C (left). All images have been convolved with a synthetic gaussain beam of width 5 AU. All simulations show clear evidence for a mm cavity, with or without radiation pressure.}\label{fig:mm_images}
\end{figure*}

\subsection{Resulting Observational properties}

We calculate the synthetic observations  for simulation A, B \& C. These simulations contain a 4.0 M$_{\rm J}$ (simulations A \& B) or 2.0 M$_{\rm J}$ (simulation C) planet orbiting at 20AU  and the accretion rate onto the star is $10^{-8}$ M$_\odot$ yr$^{-1}$. Simulation A contains no radiative feedback from the planet, while simulations B \& C do contain radiative feedback. The SEDs for simulation A (solid line), simulation B (dashed) and simulation C (dot-dashed) are shown in Figure~\ref{fig:fig3}, compared to the stellar spectrum (dotted). Simulation A (without radiative feedback) clearly shows an SED that would be classified as a primordial disc based on its profile and NIR/MIR colours. However, simulation B (with radiative feedback) shows a lack of emission at NIR wavelengths and a strong MIR emission bump. Furthermore, the 10$\mu$m silicate feature is visible indicating the presence of optically thin dust in the cavity. The SED of simulation B looks characteristically like those SEDs seen in `transition' discs and can only be created by removing the small sub-micron dust from the inner disc. Even though simulation C does include radiative feedback the planet mass (2.0 M$_J$) is not massive enough to make trapping efficient enough so the inner disc remains optically thick at NIR wavelengths giving rise to a primodial SED that is very similar to that of Simulation A.

The 880$\mu$m images are shown in Figure~\ref{fig:mm_images} for simulation A (centre) B (right) \& C (left). As expected from the mm sized particle distributions, all discs show evidence for a large cavity in the mm images. We find the peak of the mm emission occurs at roughly $1.5\times$ the planet's separation ($\sim 30$ AU) in good agreement with the model of Pinilla et al. (2012). Simulation C therefore represents a possible new observational class of `transition' disc: one which shows a primordial SED, but a large mm cavity. Two such discs (WSB 60, MWC 758) were serendipitously discovered by Andrews et al. (2011), and we will discuss the implications of this class further in Section~\ref{sec:discuss}

Therefore, in-order to reproduce both the SEDs and mm images of `transition' discs we find it is necessary to include radiative feedback from an accreting planet. However, should this radiative feedback fail, due to too low a planet mass $<3$ M$_J$ or too low a planetary accretion rate, one would expect to see observational diagnostics similar to simulation A (a mm-cavity but a primordial SED).

\section{Discussion and Perspectives}\label{sec:discuss}
Dust particles that have $\tau_s\sim 1$ - which have sizes of $\sim 1$ mm at hole radii found in `transition' discs - are trapped by pressure gradients outside the planet's orbit, consistent with the images of dust at mm wavelengths \citep{brown09,hughes09,andrews11,irs48}. Small dust particles ($\lesssim 1\,\mu$m), which dominate the opacity at IR wavelengths and ultimately whose absence give rise to the classification as a `transition' disc have $\tau_s\lesssim10^{-3}$. Our calculations and other works e.g. (Zhu et al. 2012) have shown that without radiation pressure feedback, small particles follow the gas across the gap, and give observation signatures of a primordial disc at NIR wavelengths rather than a transition disc \citep{ward09}. This was conclusively demonstrated using two-fluid simulations \citep{zhu12}, which showed that an embedded planet alone could not prevent the sub-micron dust particles entering the inner disc.  Thus, with only dust trapping due to pressure gradients, small sub-micron particles are able to cross the planet's gap entrained in the gas and give rise to an observation of a primordial disc.

Including radiation pressure from accretion onto an embedded planet allows, for the first time, the observational demographics (SED, mm image and accretion rate) of `transition' discs to be reproduced. In this model, radiation pressure holds small dust particles outside the planet's orbit, while the pressure gradient from the planet's gap traps the larger, mm size particles, such that dust depleted gas proceeds to flow into the inner disc. Ultimately, the amount of small sub-micron sized dust particles allowed to flow across the gap is set by a balance between advection, turbulent diffusion and radiation pressure, whereas for the mm sized particles it is set by the balance between turbulent diffusion and radial drift due to the pressure gradient.

It is at this point we must consider the difference between this scenario and that presented by \citet{chiang07}, who suggested radiation pressure from the star may trap dust in an MRI active layer at large radius ($\gtrsim 10$ AU). Ultimately Dominik \& Dullemond (2011), showed that (irrespective of whether you could ever set-up such a system) the dust wall was not stable, and the dust always overcame the stellar radiation pressure on a time-scale shorter than the disc's lifetime. They found that dust advection behind the radiation pressure-supported layer (whose width was set by the attenuation scale of the stellar radiation) built up a large enough concentration gradient, such that diffusion was able to overcome the radiation pressure and drive the radiation supported layer inwards. At first glance the ratio of the advective acceleration to the radiation pressure in our scenario (Equation~\ref{eqn:ratio}) is not much larger than that adopted by \citet{dominik11}. Thus, we must wonder where the difference lies. Specifically, that radiation pressure from the star is unable to hold back the dust in a standard disc, but radiation pressure from an accreting, gap-creating planet is. 

The answer lies in the fact the planet creates a deep gap reducing the surface density significantly within distances of order $(R-a)\lesssim 5R_H$, due to the planetary torque. This reduction in surface density has several effects: i) the reduction in surface density reduces the coupling between the gas and dust, and hence the advective drag force ($\propto \Sigma_g$). Meaning in reality, at the gap edge Equation~\ref{eqn:ratio} is an under-estimate (often severe) of the ratio of the radiation pressure to the advective drag force. Since this depends on the reduction in the surface density then it is obviously sensitive to the planet mass and this is one of the main causes of the planetary mass dependence found in results, rather than the increase in accretion luminosity. ii) The planet keeps the density in the gap low enough that the radiation from the planet remains optically thin until the radiation pressure and other forces become comparable, meaning the width of the radiation pressure-supported layer is not set by sharp attenuation of the radiation field, but rather by the properties of the gap (which is of a much large scale than the attenuation scale found in Dominik \& Dullemond, 2011). This allows the dust to extract more integrated momentum from the radiation field, while the increased width of the radiation-supported layer lowers the concentration gradient, weakening the diffusion that ultimately overcame the radiation field in Dominik \& Dullemond's case. iii) The torque due to the planet acts to counter-balance the advective force of the gas flow, thus the `snow-plow' formed in Dominik \& Dullemond's calculation is considerably wider, weaker and forms on a much longer time-scale, meaning turbulent diffusion is unable to overwhelm radiation pressure. Finally we note that Dominik \& Dullemond (2011) assumed a different (and somewhat unrealistic) dust distribution than that adopted in our calculation, where they took all the dust to be of a single size of $s=0.1\,\mu$m. While the opacity of an individual dust grain is similar in both calculations, the attenuation of the radiation field in the Dominik \& Dullemond (2011) calculation is much stronger then in our calculations. Our calculations use a more realistic dust distribution that has grown to $\sim$mm sizes, as observed in protoplanetary discs \citep[e.g.][]{birnstiel10,irs48,pinilla14}, rather than one dominated by small (hence high opacity) dust particles. 
Thus, in our calculations we find that, unlike Dominik \& Dullemond (2011)'s problem, we find in the planetary hypothesis for `transition' discs, radiation pressure is able to overcome turbulent diffusion and trap small dust particles at the outer edge of the gap.


\subsection{Caveats and Limitations}

In this work we have shown that radiation pressure from an accreting planet is sufficient to trap sub-micron sized dust particles outside a planet gap in a 1D simplified axis-symmetric `transition' disc model. However, since we have adopted a somewhat crude approach we need to pay special attention to the limitations of our models and discuss any possible caveats.

The high numerical overhead (in order to accurately capture the physics)  of performing a full parameter span and evolutionary calculations (where one would let the planet migrate and grow in mass) requires significant computation cost. Thus, this study aimed to look for approximate steady-state models. Therefore, we chose disc parameters to closely resemble typical `transition' discs. For example, our choice of $a=20$ AU and $\dot{M}_*=10^{-8}$ M$_\odot$ yr$^{-1}$ closely resembles GM Aur, a well known `transition' disc with an accretion rate of $\sim 10^{-8}$ M$_\odot$ yr$^{-1}$ and an inferred hole radius of $\sim$20 AU \citep{calvet05,hughes09}. Since dust particles grow and can be fragmented, in order to make detailed predictions about the final dust distributions one would need to include a dust evolution model \citep{birnstiel10,pinilla12}.  Therefore, more work needs to be done to asses which `transition' discs are created by an accreting planet, and which are driven by photoevaporation before we can draw conclusions about the planet formation process. 

Probably the major caveat of our calculations is we adopted a 1D approach, where the interaction between the planet and gas, along with the radiation field and dust, is treated in an orbit averaged sense, and a `leakage' prescription is used to model the gas flow across the planetary gap. It is well known the planet-disc interaction is a complex and highly non-axis-symmetric process; therefore, the models presented in this work can only be considered a proof of concept, rather than a fully-fledged model. {\rd Additionally \citet{fung14} have shown that radiation pressure may trigger non-axis-symmetric instabilities in accretion discs.} Only with multi-dimensional hydrodynamic models, which include dust and radiation from an accreting planet, could we confirm whether such a trapping scenario would work in practice. However, since the instantaneous radiation pressure is highest in the vicinity of where the accretion streams cross the planetary gap, this gives us hope that it will be a viable solution to the `transition' disc scenario and certainly warrants further investigation. Furthermore, the amount of gas the planet accretes from the incoming mass-flux is still rather uncertain. 
Our discussion in Section~2 indicates that for this model to work it requires the planet to accrete a significant fraction, if not the majority of the incoming gas. It still remains to be seen whether this can be envisaged in practice.   

Finally, we point out a self-consistency problem with our model as it stands, that needs to be resolved in more detailed models. Namely, we have assumed that the radiation field from the planet is dominant in the vicinity of the planet, but have not included any of the possible thermal effects this may have on the resulting gas, in particular the gap structure which is sensitive to the temperature profile of the disc and may change the response of the planet-disc interaction, and is likely to change the migration properties of the planet.

\subsection{Possible evolutionary scenario for a `transition' disc}

 Our discussion in Section~2 suggests we need a significant accretion rate onto the planet and our numerical results suggest we need a threshold planet mass. In order for a planet hosting disc to observationally appear as a `transition' disc (dip in the SED \& mm cavity), we suspect forming, embedded planets are likely to exist in several distinct observational phases. At early times and low planet masses, when the planet is unable to open and clear a gap in the gas disc (when the planet is below the thermal gap-opening limit \citealt{linpap93} $\sim 0.2$ M$_{\rm J}$), the planet will be fully embedded and the disc will appear like a primordial disc in both the SED and mm images. Once the planet grows in mass and can open a gap in the gas disc it will begin to trap mm sized dust particles giving rise to a cavity in the mm images (although as shown in Figure~\ref{fig:fig3} this hole will not be entirely devoid of emission). Small dust particles will continue to cross the gap and still appear as a primordial disc through its SED. Finally, once the planet grows to a mass of $\sim 3-4$ M$_{\rm J}$ it can trap the sub-micron grains by radiation pressure from its accretion luminosity, finally giving rise to the SED signature of a `transition disc'. Since the observed accretion rates are of order $\dot{M}\sim10^{-9}-10^{-8}$ M$_\odot$ yr$^{-1}$ in `transition' discs, and the masses required to finally appear as a `transition' disc are a few Jupiter masses, along with the time it will take for the inner disc to drain its remaining small dust particles, then this would imply the lifetime of these particular `transition' discs is relatively long-lived $\sim 0.5-1$ Myr. We caution this is not inconsistent with the observational requirement of a rapid dispersal phase (e.g. Kenyon \& Hartmann, 1995, Ercolano et al. 2012, Koperferl et al. 2013). The X-ray photoevaporation model \citep{owen10,owen11,owen12} already provides a rapid disc dispersal process and can explain a significant fraction of the observed `transition' discs (those with small holes and low accretion rates Owen \& Clarke, 2012). 

Such an evolutionary scenario would solve the `planet-mobility' problem posed by \citet{CO13}, where if one used an embedded planet to explain the `transition' discs with large holes, high accretion rates and high mm-fluxes, the planets then migrated into a forbidden region of the observable parameter space for `transition' discs. Specifically, a planet in a disc with a high mm-flux will migrate so that it gets closer to the star, giving rise to a `transition' disc with a small hole size, low accretion rate and high mm flux which is not observed \citep{OC12}. This new model would allow the planetary hypothesis for `transition' discs to solve the problem posed by \citet{CO13} by the disc simply not presenting as a `transition' disc through a dip in the SED (by allowing the small dust to refill the cavity) at small hole sizes and low accretion rates.

The scenario proposed here (an attempt to explain those `transition' discs which photoevaporation cannot), would suggest that this population of `transition' discs is a relatively rare and long lived phase of protoplanetary disc evolution, and would not actually represent a population of discs rapidly clearing from inside-out, but more excitingly a population of discs caught in the act of forming massive planets. A comparison of discs which show a `standard' transition disc signature (i.e. dip in the SED \& mm cavity) versus those which just show a mm cavity and primordial SED would allow us to observationally put constraints on the growth rate of massive planets. Currently, most discs thought to host embedded planets have been detected through SED modelling; however, two discs with a mm cavity but primordial disc SEDs have been discovered serendipitously in a mm survey \citep{andrews11} and may represent embedded planets above the thermal gap opening limit but less than 3-4 M$_{\rm J}$. A more detailed mm imaging survey of discs with primordial SEDs will be able to place constraints on the time-scale and locations of planets at the early stage of the planet-forming process. 

Additionally, the high implied accretion luminosities suggest that such accreting planets should be detectable inside the gaps of `transition' discs that present with suitably face-on inclinations. Several such candidates have been detected using IR AO imaging \citep{huelamo11,kraus12} in the `transition' discs LkCa H$\alpha$ 15 and T Cha; while these companions await confirmation (e.g. Olofsson et al. 2013) it is certainly a promising avenue to test the model. Perhaps more exciting is the discovery of an accreting planetary companion in `transition' disc HD 142527 using AO imaging at H$\alpha$ (Close et al. 2014). Close et al. (2014) use the T-Tauri star accretion luminosity - H$\alpha$ scaling to estimate an accretion Luminosity of $1.2\times10^{-2}$~L$_\odot$. This is obviously lower that that assumed in our model of $7\times10^{-2}E$~L$_\odot$; however, Close et al. (2014) do caution that using the T-Tauri star scaling at planetary masses can only represent a very rough estimate of the accretion luminosity. Close et al. (2014) argue that detection of accreting planetary mass objects in H$\alpha$ AO images of `transition' discs should be easier than at IR wavelengths, and represent a promising observational avenue to test this scenario as the model described predicts that such an accreting object should be detectable in the inner holes of all `transition' discs with high accretion rates and large holes.

{\bl Finally, once the migration of the planet-disc system is understood within the framework of this model (something not attempted here). One can compare the planetary origin for `transition' discs with the exoplanet statistics. If all the massive planets were to migrate to small separations ($\sim 1-5$ AU), then the fraction of observed `transition' discs (5-10\% \citep[e.g.][]{koepferl13}) is in slight tension with the observed fraction of massive planets from exoplanet studies \citep[e.g.][]{gaidos13}. However, it remains to be seen whether such planets can migrate to small separations (the planet itself may trigger photoevaporative disc dispersal at larger separations $\gtrsim 10$ AU - e.g. Rosotti et al. 2013). Thus, as the models and exoplanet statistics improve we maybe able to tie the exoplanet data into models of planet formation and migration.} 

\section{Summary}
In this work we have presented an update to the standard planetary hypothesis for the origin of `transition' discs by including feedback on the small dust particles from radiation pressure due to an accreting, embedded planet. Adopting the standard picture of planet-disc interaction for massive $\gtrsim $M$_J$ planets, which carve deep gaps, while allowing on-going accretion onto the star, the observed accretion rates in `transition' discs (that cannot be explained by photoevaporation) imply a high accretion luminosity originating from the forming planet $\gtrsim 10^{-3}$ L$_\odot$. At the gap edge, radiation pressure from the planetary accretion luminosity can be the dominant force on small ($\lesssim 1\,\mu$m) dust particles. 

Using a simple 1D secular coupled gas and dust model for the disc, we find, in agreement with previous studies, that without radiative feedback the planet cannot explain all the observed features of `transition' discs. Without radiative feedback, massive planets can open deep enough gaps to trap the mm sized dust particles, while allowing the small dust particles to follow the gas into the inner disc, giving rise to a primordial SED but mm cavity. However, including radiative feedback from the accreting planet we find that above a planet mass threshold of $\sim 3-4$ M$_J$ radiation pressure is able to hold back the small $s\lesssim 1\,\mu$m dust particles, allowing dust free gas to accrete across the gap and into the inner disc. We require that the planets accrete at least half of the material flowing into the gap in-order to trap sufficient sub-micron dust particles. Computing synthetic SEDs and mm images we find we are able to explain the observed NIR dip, mm cavity and accretion rate within a single model.  

The fact this process possesses a planetary mass threshold suggests that `transition' discs with large holes and high accretion rates are not in-fact discs rapidly transitioning from a primordial to disc-less state, but rather a rare and relatively long-lived state (0.5-1 Myr). Assuming a scenario where planets accrete and grow above the required mass-threshold to appear as a `transition' disc, allows us to construct an evolutionary scenario where a planet-hosting disc would first show a mm cavity but primordial SED while the planet mass is low. Once it grows above the mass threshold it presents as a `transition' disc with mm cavity and dip in the SED at NIR wavelengths. Finally, once the accretion rate in the disc (and hence onto the star and planet) drops or onto the planet the small dust then refills the inner cavity again giving rise to a disc that has a mm cavity but primordial SED again, before photoevaporation finally disperses the disc.

Finally this model suggests that `transition' discs with large holes and high accretion rates should all have heavily accreting planetary objects. These planets would be detectable at close to face-on inclinations using AO imaging at H$\alpha$, as demonstrated by the detection of an accreting planetary object in the `transition' disc HD 142527 using the Magellan Adaptive Optics VisAO camera.

\acknowledgments
JEO thanks the referee for comments which improved the manuscript. JEO is grateful to Cathie Clarke, Ruobing Dong, Barbara Ercolano, Giovanni Rosotti, Ilaria Pascucci, Yanqin Wu and Zhaohuan Zhu for helpful discussions. The calculations were performed
on the Sunnyvale cluster at CITA which is funded by the
Canada Foundation for Innovation.

\bibliographystyle{apj}
\bibliography{bib_paper}

\appendix

\section{Evaluating the orbit averaged radiation pressure term}

\begin{figure*}
\centering
\includegraphics[width=0.5\textwidth]{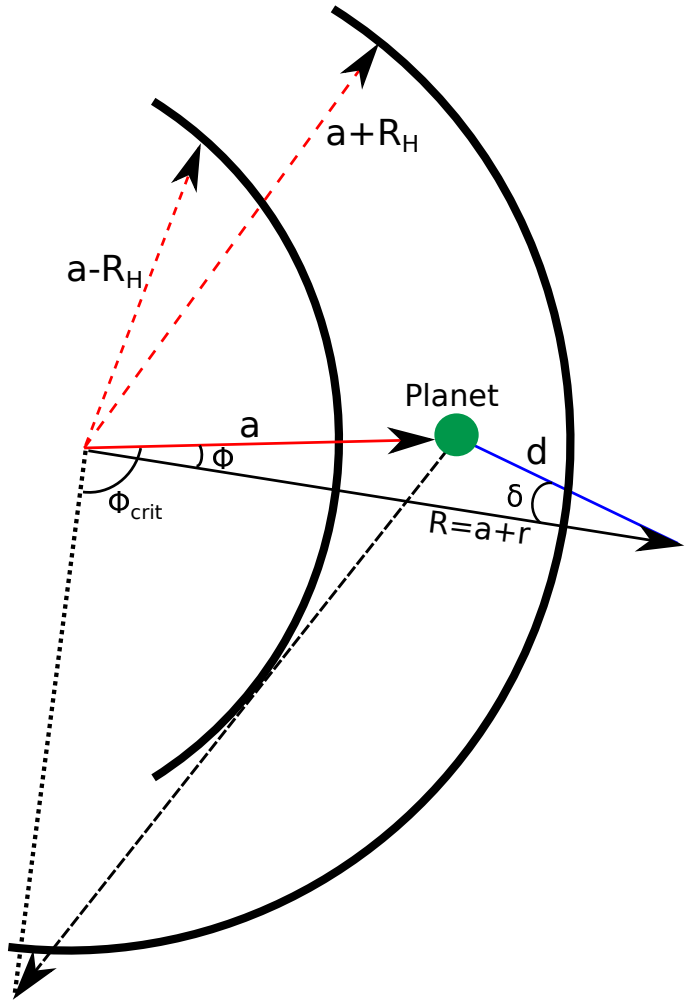}
\caption{Schematic diagram of the geometric setup, the thick lines show the gap edges at $\pm R_H$ from the planet at a separation $a$. At a given point $\{R,\phi\}$, $d$ represents the distance to the planet, $\delta$ is the angle between the planet-point vector and the radial unit vector. The angle $\phi_{\rm crit}$ indicates the maximum angle that has direct line of sight to the planet at a given $R$, before it is blocked by the gap edges. }
\label{fig:fig1}
\end{figure*}

The instantaneous acceleration on a dust particle due to radiation pressure is given by:
\begin{equation}
{\bf a}^{\rm rad}=\frac{\kappa{\bf F}^p}{c}
\end{equation}
However, since we are interested in the long term evolution of small dust particles, which are tightly coupled to the gas ($\tau_s\ll 1$) they will move slowly through a succession of Keplerian circular orbits \citep{takeuchi02}. Therefore, we can consider this slow secular evolution by averaging the radiation pressure over an entire orbit. We describe the geometry of this setup in Figure~\ref{fig:fig1}, {\bl where we have transformed into a frame co-rotating with the planet}.
Considering an orbit averaged radiation pressure:
\begin{eqnarray}
\left<a^{rad}_R\right>&=&\frac{\frac{\kappa}{c}\int_0^TF^p_R{\rm d}t}{T}\\
\left<a^{rad}_\phi\right> & = &\frac{\frac{\kappa}{c}\int_0^TF^p_\phi{\rm d}t}{T}=0
\end{eqnarray}
where $T$ is the orbital period {\bl at a radius $R$ in the co-rotating frame, i.e. $T=2\pi[\Omega(R)^{-1}-\Omega_p(a)^{-1}]=2\pi/\tilde{\Omega}$}. Clearly the orbit azimuthal component is zero by symmetry.
The radial component of the planet's flux in the radial direction is given by:
\begin{equation}
F^p_R=\frac{L_p\exp(-\frac{\tau_R}{\cos\delta})}{4\pi d^2}\cos\delta
\end{equation}
where $\tau_R$ is the radial (mid-plane) optical depth from the planet to a cylindrical radius $R$ (i.e. $\tau_R=\int_a^R\kappa\rho{\rm d}R$) and we have used attenuation in slab geometry to include optical depth effects (valid provided $d<a$). One can find an analytic solution in the optically thin limit by proceeding as follows: defining $r=R-a$ as the cylindrical radial distance between the disc material at radius $R$ and the planet, and $d$ is the distance between the planet and a region of the disc at a given ${R,\phi}$ position, then using the rule of cosines one can show:
\begin{equation}
d=\sqrt{a^2+(a+r)^2-2a(a+r)\cos\phi}
\end{equation}
and
\begin{equation}
\cos\delta=\frac{a+r-a\cos\phi}{  \sqrt{a^2+(a+r)^2-2a(a+r)\cos\phi}}
\end{equation}
Thus, the equation for the orbit averaged acceleration due to radiation pressure is:
\begin{equation}
\left<a^{\rm rad}_R\right>=\frac{2\kappa L_p}{4\pi c a^2 T}\int_0^{t_{\rm crit}}\frac{1+\tilde{r} -\cos\phi}{\left[1+(1+\tilde{r})^2-2(1+\tilde{r})\cos\phi\right]^{3/2}}{\rm d}t
\end{equation}
where $\tilde{r}=r/a$. Now, replacing ${\rm d}t$ with {\bl ${\rm d}\phi/\tilde{\Omega}$} and using {\bl $T=2\pi/\tilde{\Omega}^{-1}$} then the equation simply becomes:
\begin{equation}
\left<a^{\rm rad}_R\right>=\frac{2\kappa L_p}{8\pi^2 c a^2}\int_0^{\phi_{\rm crit}}\frac{1+\tilde{r} -\cos\phi}{\left[1+(1+\tilde{r})^2-2(1+\tilde{r})\cos\phi\right]^{3/2}}{\rm d}\phi
\end{equation}
where $\phi_{\rm crit}$ is given by:
\begin{equation}
\phi_{\rm crit}={\rm arcos}\left(1-\frac{R_H}{a}\right)+{\rm arcos}\left(\frac{1-R_H/a}{1+r/a}\right)
\end{equation}
the integral can be evaluated and formally has the solution:
\begin{eqnarray}
\left<a^{\rm rad}_R\right>&=&\frac{2\kappa L_p}{8\pi^2 c a^2}(\tilde{r}^2 \sqrt{(2 + 2\tilde{r} + \tilde{r}^2 - 2(1 + \tilde{r}) \cos\phi_{\rm crit})/\tilde{r}^2} \,E(\phi_{\rm crit}/2, -2\sqrt{1 + \tilde{r}})/\tilde{r})\nonumber\\ &+& \tilde{r} (2 + \tilde{r}) \sqrt{(2 + 2\tilde{r} + \tilde{r}^2 - 2(1 + \tilde{r}) \cos\phi_{\rm crit})/\tilde{r}^2} F(\phi_{\rm crit}/2, -2\sqrt{1 + \tilde{r}}/\tilde{r})\nonumber\\ &+& 2(1 + \tilde{r}) \sin\phi_{\rm crit})/(\tilde{r} (1 + \tilde{r}) (2 + \tilde{r}) \sqrt{2 + 2 \tilde{r} + \tilde{r}^2 - 2 (1 + \tilde{r}) \cos\phi_{\rm crit}})
\end{eqnarray}
where $E(\theta,a)$ \& $F(\theta,a)$ are elliptic integral of the first and second kind. What is not immediately obvious from this expression is that the result is effectively independent of $\phi_{\rm crit}$ and thus our choice of where to place the gap edge. This arises since most of the impulse comes when the dust particle is close to the planet in its orbit, rather than at large angles. In the numerical calculations we include the attenuation of the radiation due to absorption by the dust, and thus  evaluate the orbit averaging integral numerically. 

\section{Numerical tests}\label{App:code_test}
\begin{figure}[ht]
\centering
\includegraphics[width=0.6\textwidth]{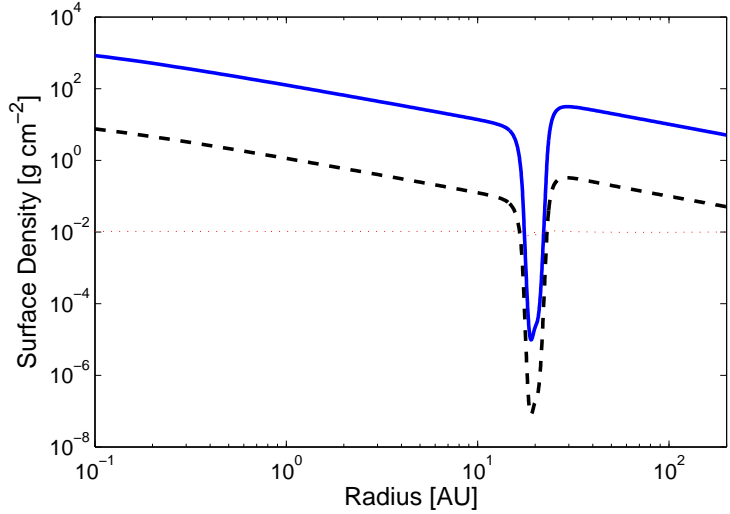}
\caption{Dust and gas surface densities for passive scalar test. The solid line shows the gas, the dashed shows the small ($s=1\times10^{-50}$ cm dust particles and the dotted shows the dust to gas ratio. This calculation using a planet mass of 3 M$_J$ an accretion rate of $10^{-8}$ M$_\odot$ yr$^{-1}$ and the standard value of $E=6$ is used. The grid is as described in Section~3}
\label{fig:test}
\end{figure}

 We test the numerical scheme is behaving as expected by considering several test problems. First to test the `leakage' implementation across the gap does not provide artificial dust trapping we consider a  simple but highly relevant test problem. A very small dust particle ($s=10^{-50}$ cm for the test), which will behaviour as a passive scalar should exactly follow the gas distribution, as at such a small size it does not feel dust drag. Therefore, in the absence of radiation pressure the dust-to-gas mass ratio should be radially fixed for the entire simulation range (excluding the region inside the planet hill sphere where the source and sink terms dominate) and the dust particles should be perfectly advected across the gap by the `leakage' scheme. The result of this test calculation is shown in Figure~\ref{fig:test}, where we find our scheme behaves accurately, and is suitable for the required calculation. We also perform a resolution study and check our scheme does conserve mass to machine precision as expected. 
 
\begin{figure*}
\centering
\includegraphics[width=\textwidth]{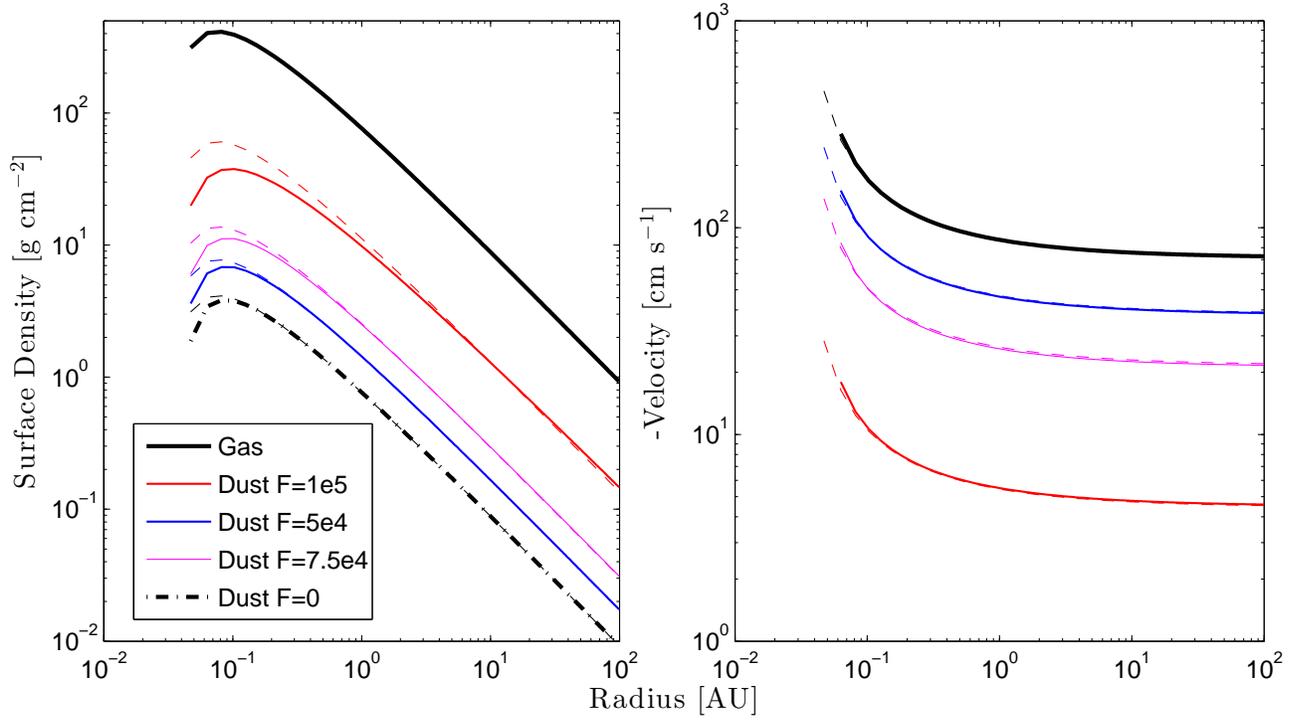}
\caption{Figure showing the comparison between the numerical solutions (solid) and analytic solutions (dashed) to the steady-state problem with radiation pressure described in the text. The left panel shows the surface density, and the right pannel shows the negative velocity. In both panels the thick solid lines shows the gas properties. We note we only expect good agreement far from the inner boundary, which cannot be captured accurately analytically, or numerically.}\label{fig:arad_test}
\end{figure*}

Furthermore, we can test our that we are accurately capturing the effect of radiation pressure on the small dust particles. We can construct a steady-state approximate analytic solution to the dust and gas problem. Assuming the dust particles are well coupled $\tau_s\ll 1$, such that we can safely neglect dust drag then the radial velocity of the dust particles is approximately given by:
\begin{equation}
v_R\approx u_R+\frac{\tau_s\kappa F^p_R}{c\Omega}
\end{equation}
If we consider a steady disc, with constant influx of dust and gas at the outer boundary then we may write:
\begin{equation}
v_R=-\left[\frac{\dot{M_*}}{2\pi R \Sigma_g}-\left(\frac{3\pi}{8c\Sigma_g\Omega}\right)F^p_R\right]
\end{equation}
Given for a steady disc we can write the surface density of the gas as (e.g. Pringle 1981):
\begin{equation}
\Sigma_g=\frac{\dot{M}_*}{3\pi\nu}\left(1-\sqrt{\frac{R_{\rm in}}{R}}\right)\label{eqn:sigma_steady}
\end{equation}
where $R_{\rm in}$ is the inner radius of the disc. Noticing for $\nu\propto R$, if $F^p_R\propto R^{-2.5}$ then the dust and gas velocity have the same radial dependence (specifically the are constant with radius of $\nu\propto R$), then for $R\gg R_{\rm in}$ the dust concentration will be constant. Therefore, we can ignore the dust diffusion term and simply find the dust surface density as:
\begin{equation}
\Sigma_d=\frac{X\dot{M}_*}{2\pi R \left[\frac{\dot{M_*}}{2\pi R \Sigma_g}-\left(\frac{3\pi}{8c\Sigma_g\Omega}\right)F^p_R\right]}
\end{equation}
where $X$ is the dust-to-gas mass ratio of the material injected into the disc at the outer boundary and $\Sigma_g$ is given by Equation~\ref{eqn:sigma_steady}. We simulate the this steady problem with an accretion rate in the disc of $10^{-8}$ M$_\odot$ yr$^{-1}$, an outer boundary of 100AU, inner boundary of 0.04 AU, viscous $\alpha$ of 0.01 and $H/R$ of 0.04 at 1 AU, where $\nu\propto R$ throughout the entire grid. We use a dust particle of $\rho_d=2$ g cm$^{-3}$, radius of 0.1$\mu$m and dust-to-gas mass ratio of 0.01. We take the flux to scale as $F^p_R\propto R^{-2.5}$ and choose the Flux (measured at 10 AU) to have values of 0, $5\times10^{4}$, $7.5\times10^{4}$ and $1\times 10^{5}$ erg s$^{-2}$ cm$^{-2}$. The comparison between the analytic solution (dashed lines) for the dust surface density and velocity and simulations (solid lines) are shown in Figure \ref{fig:arad_test}. We find excellent agreement between the code and analytic solutions, indicating the radiation pressure routine in our numerical method is behaving as expected.

\end{document}